\documentclass[preprint,showpacs,preprintnumbers,amsmath,amssymb]{revtex4}

\usepackage{graphicx}
\usepackage{dcolumn}
\usepackage{bm}

\begin{document}

\preprint{}

\title{Dyson-Index-Like Behavior of Bures Separability Functions}

\author{Paul B. Slater}%
\email{slater@kitp.ucsb.edu}
\affiliation{%
ISBER, University of California, Santa Barbara, CA 93106\\
}%
\date{\today}

\begin{abstract}
We conduct a study based on the Bures 
(minimal monotone) 
metric, analogous to that 
recently reported for the Hilbert-Schmidt (flat or Euclidean) 
metric (arXiv:0704.3723v2). 
Among the interesting results obtained there
had been  
proportionalities---in {\it exact} correspondence to the 
Dyson indices $\beta= 1, 2, 4$ of random matrix theory---between the fourth, 
second and first powers of the separability functions
$\mathcal{S}_{type}(\mu)$ for real, 
complex and quaternionic  qubit-qubit scenarios, Here $\mu=\sqrt{\frac{\rho_{11} \rho_{44}}{\rho_{22} \rho_{33}}}$, with 
$\rho$ being a $4 \times 4$ 
density matrix.
Separability functions have proved 
useful---in the framework 
of the Bloore (correlation coefficient/off-diagonal scaling) 
parameterization of density matrices---for 
the calculation of separability {\it probabilities}.
We find---for certain, basic simple scenarios
(in which the diagonal entries of $\rho$ are 
unrestricted, and one or two off-diagonal
[real, complex or quaternionic] pairs of entries are nonzero) ---that these 
proportionalities no longer strictly hold in the Bures case, but do
come remarkably close to holding.
\end{abstract}


\maketitle
In our recent study \cite{slater833}, we reported a number of developments of value in resolving the clearly challenging and conceptually important 
question of the probabilities---using the measure induced by the
Hilbert-Schmdit (HS) metric---that generic real, complex 
and quaternionic 
qubit-qubit and qubit-qutrit states are separable/disentangled.
An essential component in this progress was the use of the (simple) 
parameterization of the $n \times n$ density matrices ($\rho$) that had been
originally 
proposed by Bloore \cite{bloore}.  
This involves reparameterizing the  
off-diagonal entries  
$\rho_{ij}$ as $\sqrt{\rho_{ii} \rho_{jj}} (x_{ij} + {\bf{i}}  y_{ij})$. 
In the real case ($y_{ij}=0$), the $n \times n$ matrix of 
$x_{ij}$'s, being  necessarily 
nonnegative definite, with $x_{ij} \in [-1,1]$, 
has the form of a {\it correlation matrix}---a basic
object of study in descriptive statistics \cite{joe,kurowicka,kurowicka2}.
(Correlation matrices can be obtained by standardizing {\it covariance} 
matrices. Density matrices have been viewed as covariance matrices of
multivariate normal [Gaussian] distributions \cite{guiasu}. The possible 
states of polarization of a two-photon system are describable by six
Stokes parameters and a $3 \times 3$ ``polarization correlation'' matrix 
\cite{vanik}.)

A major virtue (of course, unrecognized more than thirty years ago 
in the 1976 
paper \cite{bloore})  of the
 Bloore parameterization is that it allows one to express the 
well-known Peres-Horodecki
positive-partial-transposition criterion \cite{asher,michal,bruss} 
for separability using fewer variables
than one would naively anticipate. Since we are largely concerned with the 
evaluation of 
high-dimensional
integrals, this reduction in number of relevant variables is certainly of 
considerable importance.

Here, we parallel the sequential approach of
\.Zyczkowski and Sommers in that they, first,  computed the {\it 
total} volume of
(separable {\it and} nonseparable) $n \times n$ density matrices in terms of
the (flat or Euclidean) 
Hilbert-Schmidt metric 
\cite{szHS} \cite[secs. 9.6-9.6, 14.3]{ingemarkarol}, and then,
using the 
fundamentally important 
Bures (minimal monotone) metric \cite[sec. 14.4]{ingemarkarol} 
 \cite{szBures}.
(In particular, they employed the Laguerre ensemble of random matrix theory 
\cite{random} 
in both sets of computations (cf. \cite{andai}). The Bures and HS metrics
were compared by Hall \cite{hall}, who concluded that the Bures induced
the ``minimal-knowledge ensemble'' (cf. \cite{slatersrednicki}).)
That is, we will seek now
to extend the form of  analysis applied in the Hilbert-Schmidt context in
\cite{slater833} to the Bures setting.

To begin, let us review the most elementary findings reported in 
\cite[sec. II.A.1]{slater833}. 
The simplest (four-parameter) scenario studied there posits a 
$4 \times 4$ density matrix $\rho$ with
fully general diagonal entries ($\rho_{11}, \rho_{22}, \rho_{33}, 
\rho_{44} = 1 -\rho_{11}-\rho_{22}-\rho_{33}$) and only one pair of real off-diagonal non-zero entries,
$\rho_{23}=\rho_{32}$. The HS separability function 
for that scenario was found to take 
the form
\cite[eq. (20)]{slater833},
\begin{equation} \label{equationA}
\mathcal{S}^{HS}_{[(2,3)]}(\mu) =
\begin{cases}
 2 \mu & 0\leq \mu \leq 1 \\
 2 & \mu >1
\end{cases},
\end{equation}
where we will now primarily employ (purely as a matter of convenience) 
the variable $\mu=
\sqrt{\frac{\rho_{11} \rho_{44}}{\rho_{22} \rho_{33}}}$, rather than 
$\nu=\mu^2$, as in \cite{slater833,slaterPRA2}.

Allowing the 23- and 32-entries to be complex conjugates of one another,
we further found for the corresponding 
separability function \cite[eq. (22)]{slater833}---where the 
wide tilde over an ${i,j}$ pair will 
throughout indicate a complex entry (described by  
{\it two} parameters)---
\begin{equation} \label{equationB}
\mathcal{S}^{HS}_{[\widetilde{(2,3)}]}(\mu)=
(\sqrt{\frac{\pi}{2}} \mathcal{S}^{HS}_{[(2,3)]}(\mu))^2 =
\begin{cases}
 \pi  \mu^2  & 0\leq \mu \leq 1 \\
 \pi  & \mu >1
\end{cases}.
\end{equation}

Further, permitting 
the 23- and 32-entries to be {\it quaternionic}  conjugates of one another 
\cite{adler,asorey}, the corresponding
separability function \cite[eq. (24)]{slater833}---where the
wide hat over an ${i,j}$ pair will throughout indicate a quaternionic 
entry (described by
{\it four} parameters)---took the form
\begin{equation} \label{equationQuat}
\mathcal{S}_{[\widehat{(2,3)}]}(\nu)= 
(\sqrt{\frac{\pi}{2}} \mathcal{S}^{HS}_{[\widetilde{(2,3)}]}(\mu))^2 =
(\sqrt{\frac{\pi}{2}} \mathcal{S}^{HS}_{[(2,3)]}(\mu))^4=
\begin{cases}
 \frac{\pi^2  \mu^4}{2}  & 0\leq \mu \leq 1 \\
 \frac{\pi^2}{2}  & \mu >1
\end{cases}.
\end{equation}

So, the real (\ref{equationA}), complex (\ref{equationB}), 
and quaternionic (\ref{equationQuat})
HS separability functions accord perfectly
with the Dyson index sequence  $\beta= 1, 2, 4 $ of random matrix theory
\cite{dyson}. ``The value of $\beta$ is given by the number of independent
degrees of freedom per matrix element and is determined by the antiunitary 
symmetries \ldots It is a concept that originated in Random Matrix Theory 
and is important for the Cartan classification of symmetric spaces'' 
\cite[p. 480]{kogut}. The Dyson index corresponds to the multiplicity of 
ordinary roots, in the terminology of symmetric spaces \cite[Table 2]{caselle}. 
However, we remain unaware of any specific line of argument using random
matrix theory \cite{random} that can be used to formally confirm
the HS separability function Dyson-index-sequence 
phenomena we have noted above and observed in 
\cite{slater833}. (The basic difficulty/novelty 
appears to be that the separability
aspect of the problem introduces a totally new set of complicated
constraints---{\it quartic} (biquadratic) in $\mu$ 
\cite[eq. (5)]{slater833} \cite[eq. (7)]{slaterPRA2}---that the multivariate integration must respect 
\cite[sec. I.C]{slater833}.)

As a further recent exercise, unreported in \cite{slater833}, 
we found that
setting any single one of the four components of the quaternionic 
entry, $x_{23} +{\bf{i}}  y_{23} +{\bf{j}} j u_{23} +{\bf{k}} 
v_{23}$, in the
scenario just described, to zero, yields the separability function,
\begin{equation} \label{missing1}
\mathcal{S}^{HS}_{[\hat{(2,3)}]} =
\begin{cases}
 \frac{4 \pi  \mu ^3}{3} & 0\leq \mu \leq 1 \\
 \frac{4 \pi }{3} & \mu >1
\end{cases},
\end{equation}
consistent, at least, in terms 
of the exponent of $\mu$, with the Dyson-index 
pattern previously observed.

Continuing the analysis in \cite{slater833}, we computed 
the integrals 
\begin{equation} \label{Vsmall}
V^{HS}_{sep/scenario}= \int_{0}^{\infty} \mathcal{S}^{HS}_{scenario}(\mu) 
\mathcal{J}^{HS}_{scenario}(\mu)
d \mu,
\end{equation}
of the products of 
these separability functions with the corresponding (univariate) 
marginal jacobian functions
 (which are obtained by integration over
diagonal parameters only  and {\it not} any of 
the off-diagonal $x_{ij}$'s and 
$y_{ij}$'s) for the reparameterization of $\rho$ using the Bloore variables
\cite[eq. (17)]{slater833}. This 
yielded the HS scenario-specific {\it separable} volumes 
$V^{HS}_{sep/scenario}$. The ratios of such separable volumes  to
the HS total volumes
\begin{equation} \label{Vbig}
V^{HS}_{tot/scenario}= c_{scenario}^{HS} 
\int_{0}^{\infty}  \mathcal{J}_{scenario}^{HS}(\mu) d \mu,
\end{equation}
where $c^{HS}_{scenario}$ is a scenario-specific constant, gave 
us in \cite{slater833} (invariably, it seems, exact) separability {\it probabilities}. (For the three scenarios listed above, these probabilities were, 
respectively, $\frac{3 \pi}{16}, \frac{1}{3}$ and $\frac{1}{10}$.)

Based on the numerous scenario-specific 
analyses in \cite{slater833}, we are led to believe
that the real, complex and quaternionic separability functions adhere to
the Dyson-index pattern for general scenarios, when
the Hilbert-Schmidt measure has been employed. This apparent adherence
was of central importance 
in arriving at the assertions in \cite[secs.~IX.A.1 and 
IX.A.2]{slater833} that the HS separability probabilities of generic 
[9-dimensional] real
and [15-dimensional] 
complex two-qubit states 
are $\frac{8}{17}$ and 
$\frac{8}{33}$, respectively. 
There we had posited---using mutually supporting numerical and theoretical 
arguments---that \cite[eq. (102)]{slater833}
\begin{equation}
\mathcal{S}_{real}(\mu) \propto  \frac{1}{2} (3- \mu^2) \mu,
\end{equation}
and, further pursuing our basic Dyson-index ans{\"a}tz (fitting our 
numerical simulation extremely well \cite[Fig. 4]{slater833}), that
$(\mathcal{S}_{real}(\mu))^2 \propto \mathcal{S}_{complex}(\mu)$.
(Obviously, we must as well make the further claim 
 that $(\mathcal{S}_{real}(\mu))^4 
\propto \mathcal{S}_{quat}(\mu)$. Unlike the real and complex
cases, however, we have performed no numerical analyses 
such as those in \cite{slaterPRA2} to guide us as to the proper 
coefficient of proportionality to employ. Thus, we have no 
specific assertion to advance 
as to the two-qubit quaternionic separability probability---although
$\frac{8}{65}$ or $\frac{8}{129}$ might be readily suggested.)

Now, employing 
formulas (13) and (14) 
of Dittmann \cite{explicit} for the {\it Bures} metric---which 
avoid the possibly problematical need for diagonalization 
of $\rho$---we were able to find the
Bures volume elements for the same three 
basic (one pair of free off-diagonal entries) 
scenarios. We obtained for the real case,
\begin{equation} \label{V23real}
dV^{Bures}_{[(2,3)]} = \frac{\sqrt{\rho _{11}} \sqrt{1-\rho _{11}-\rho _{22}}
   \sqrt{\rho _{22}}}{4 \sqrt{1-x_{23}^2} \left(\rho _{22}
   \mu ^2+\rho _{11}\right) \sqrt{\mu ^2 \rho
   _{22}^2+\left(1-\rho _{11}\right) \rho _{11}}} d \rho_{11} d\rho_{22} d x_{23} d \mu,
\end{equation}
for the complex case,
\begin{equation} \label{V23complex}
dV^{Bures}_{[\widetilde{(2,3)}]}= \frac{\rho _{11} \rho _{22} \left(\rho _{11}+\rho
   _{22}-1\right)}{4 \sqrt{1-y_{23}^2-x_{23}^2}
   \left(\rho _{22} \mu ^2+\rho _{11}\right) \left(-\rho
   _{11}^2+\rho _{11}+\mu ^2 \rho _{22}^2\right)} 
d \rho_{11} d\rho_{22} d x_{23} d y_{23} d \mu,
\end{equation}
and for the quaternionic case,
\begin{equation} \label{V23quaternionic}
dV^{Bures}_{[\widehat{(2,3)}]}= \frac{A}{B} 
d \rho_{11} d\rho_{22} d x_{23} d y_{23}  d u_{23} d v_{23} d \mu,
\end{equation}
where
\begin{displaymath}
A=  -\rho _{11}^2 \rho _{22}^2 \left(\rho _{11}+\rho
   _{22}-1\right)^2,
\end{displaymath}
and
\begin{displaymath}
B=4 \sqrt{1-u_{23}^2-v_{23}^2-x_{23}^2-y_{23}^2} \left(\rho
   _{22} \mu ^2+\rho _{11}\right) \left(-\rho _{11}^2+\rho
   _{11}+\mu ^2 \rho _{22}^2\right)^2.
\end{displaymath}

In analyzing the
quaternionic case, we transformed---using standard 
procedures \cite[p. 495]{adler} \cite[eq. (17)]{slaterJMP1996}---the 
corresponding $4 \times 4$ density matrix into an $8 \times 8$ density matrix with (only) complex entries. To
this, we found it most convenient to apply---since its 
eigenvalues and eigenvectors 
could be explicitly computed---the basic formula of H\"ubner 
\cite{hubner} \cite[p. 2664]{dittmann} for the Bures metric. 

Integrating these three volume elements  over all the 
(four, five or seven) 
variables, while enforcing nonnegative definiteness of $\rho$, we derived
the Bures 
{\it total} (separable {\it and} nonseparable) volumes for the three
scenarios---$V^{Bures}_{tot/[(2,3)]} =\frac{\pi^2}{12} 
\approx 0.822467$, 
$V^{Bures}_{tot/[\widetilde{(2,3)}]}=\frac{\pi^3}{64} \approx 0.484473$, and 
$V^{Bures}_{tot/[\widehat{(2,3)}]}=\frac{\pi^4}{768} \approx 0.126835$.

We note importantly that
the Bures volume elements ((\ref{V23real}), (\ref{V23complex}), 
(\ref{V23quaternionic})), in these three cases, can be 
{\it factored} into products of functions of 
the off-diagonal variables, $u_{23}, v_{23}, x_{23}$ and $y_{23}$, 
and functions of the diagonal 
variables, $\rho_{11}, \rho_{22}$ and $\mu$. Now, we will integrate 
(one may transform to polar and spherical coordinates, as appropriate) 
just the 
factors ---$\frac{1}{\sqrt{1-x_{23}^2}}$, 
$\frac{1}{\sqrt{1-x_{23}^2 - y_{23}^2}}$ 
and $\frac{1}{\sqrt{1-u_{23}^3-v_{23}^2-x_{23}^2-y_{23}^2}}$---involving
the off-diagonal variable(s) over 
those variables. In doing this, we will further enforce 
(using the recently-incorporated 
integration-over-implicitly-defined-regions feature of Mathematica)
the Peres-Horodecki positive-partial-transpose-criterion
\cite{asher,michal,bruss}, 
expressible as
\begin{equation}
\mu^2 -x_{23}^2 \geq 0
\end{equation}
 in the real case,
\begin{equation}
\mu^2 -x_{23}^2 -y_{23}^2 \geq 0,
\end{equation}
 in the complex case, and 
\begin{equation}
\mu^2 -x_{23}^2 -y_{23}^2 - u_{23}^2 -v_{23}^2 \geq 0,
\end{equation}
in the quaternionic case. (None of the individual diagonal $\rho_{ii}$'s 
appears explicitly in these constraints, due to an attractive property of the
Bloore [correlation coefficient/off-diagonal scaling] 
parameterization. Replacing $\mu^2$ in these three constraints by 
simply unity, we obtain the non-negative definiteness constraints on $\rho$ itself, which we also obviously must enforce.) 
Performing the indicated three integrations, we obtain
the {\it Bures} separability functions,
\begin{equation} \label{Bures1}
\mathcal{S}^{Bures}_{[(2,3)]}(\mu) = 
\begin{cases}
 \pi  & \mu \geq 1 \\
 2 \sin ^{-1}(\mu ) & 0 < \mu <1
\end{cases},
\end{equation}
\begin{equation} \label{Bures2}
\mathcal{S}^{Bures}_{[\widetilde{(2,3)}]}(\mu) = 
\begin{cases}
 2 \pi  & \mu \geq 1 \\
 2 \pi  \left(1-\sqrt{1-\mu ^2}\right) & 0<\mu <1
\end{cases},
\end{equation}
and
\begin{equation} \label{Bures3}
\mathcal{S}^{Bures}_{[\widehat{(2,3)}]}(\mu) =
\begin{cases}
 \frac{4 \pi ^2}{3} & \mu >1 \\
 \frac{2}{3} \pi ^2 \left(-\sqrt{1-\mu ^2} \mu ^2-2
   \sqrt{1-\mu ^2}+2\right) & 0 <\mu <1
\end{cases}.
\end{equation}
Then, utilizing these three separability functions---that is, 
integrating the products of the functions and the 
corresponding remaining 
{\it diagonal}-variable factors 
in the Bures volume elements ((\ref{V23real}), (\ref{V23complex})), 
((\ref{V23quaternionic})) 
over the $\mu, \rho_{11}$ and $\rho_{22}$ variables---we obtain
{\it separable} volumes of $V^{Bures}_{sep/[(2,3)]}= 0.3658435525$ and
\begin{equation}
V^{Bures}_{sep/[\widetilde{(2,3)}]}=  
V^{Bures}_{tot/[\widetilde{(2,3)}]} - \frac{1}{32} \pi ^2 (-2 C+\pi ) = 
\frac{1}{64} \pi ^2 (4 C-6 +\pi ) 
\approx 0.124211 
\end{equation}
 and consequent
separability {\it probabilities}, respectively, 
of 0.4448124200 and (our only {\it exact} Bures separability probability
result in this study (cf. \cite{slaterC})),
\begin{equation} \label{exactprob}
P^{Bures}_{sep/[\widetilde{(2,3)}]} = \frac{4 C-6+\pi }{\pi } 
\approx 0.256384,
\end{equation}
where $C \approx 0.915966$ is Catalan's constant (cf. \cite{collins}). 
(This constant appears
commonly in estimates of 
combinatorial functions and in certain classes of sums and definite 
integrals \cite[sec.~1.7]{finch}.) Further, for the quaternionic case, 
$V^{Bures}_{sep/\widehat{[(2,3)]}} \approx 0.012954754466$, and 
$P^{Bures}_{sep/\widehat{[(2,3)]}} \approx  0.10213883862$.
 (The corresponding HS separability probability was also of the 
same relatively 
small magnitude, that is, $\frac{1}{10}$ \cite[sec.~II.A.3]{slater833}.
We have computed the various Bures separable volumes and probabilities
to high numerical accuracy, hoping that such accuracy may be useful
in searches for possible further exact formulas for them.)

\begin{figure}
\includegraphics{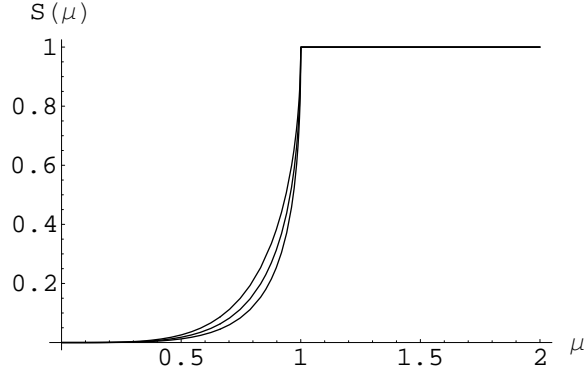}
\caption{\label{fig:functs} Joint plot of the 
normalized Bures {\it quaternionic} 
separability function 
$\frac{3 \mathcal{S}^{Bures}_{[\widehat{(2,3)}]}(\mu)}{4 \pi^2}$, 
the {\it square} of 
the normalized Bures {\it complex} separability function
$\frac{\mathcal{S}^{Bures}_{[\widetilde{(2,3)}]}(\mu)}{2 \pi}$,
and the {\it fourth} power of the normalized Bures {\it real} separability
function $\frac{\mathcal{S}^{Bures}_{[(2,3)]}(\mu)}{\pi}$. The order of
dominance of the three curves is the same as the order in which they have been
mentioned.}
\end{figure}

So, the normalized---to equal 1 at $\mu=1$---forms of these three
separability 
functions are $\frac{\mathcal{S}^{Bures}_{[(2,3)]}(\mu)}{\pi}$,
$\frac{\mathcal{S}^{Bures}_{[\widetilde{(2,3)}]}(\mu)}{2 \pi}$ and 
$\frac{3 \mathcal{S}^{Bures}_{[\widehat{(2,3)}]}(\mu)}{4 \pi^2}$. 
In Fig.~\ref{fig:functs}, we plot---motivated by the appearance of the 
Dyson indices in the analyses of 
\cite{slater833}---the {\it fourth} power of the first (real) of these
three normalized functions together with the {\it square} of the 
second (complex) function and the (untransformed) third 
(quaternionic) function itself. 
We find a very close, 
\begin{equation}
\Big(\frac{\mathcal{S}^{Bures}_{[(2,3)]}(\mu)}{\pi}\Big)^4  \approx (\frac{\mathcal{S}^{Bures}_{[\widetilde{(2,3)}]}(\mu)}{2 \pi})^2 
\approx (\frac{3 \mathcal{S}^{Bures}_{[\widehat{(2,3)}]}(\mu)}{4 \pi^2}),
\end{equation}
but now {\it not} exact fit, as we did find 
in \cite{slater833}  for their (also normalized) 
Hilbert-Schmidt
counterparts  $\frac{\mathcal{S}^{HS}_{[(2,3)]}(\mu)}{2}$,
$\frac{\mathcal{S}^{HS}_{[\widetilde{(2,3)}]}(\mu)}{\pi}$ 
and $\frac{2 \mathcal{S}^{HS}_{[\widehat{(2,3)}]}(\mu)}{\pi^2}$ ((\ref{equationA}), (\ref{equationB}), (\ref{equationQuat})).

As an additional exercise (cf. (\ref{missing1})), 
we have computed the Bures separability function
in the case that a single one of the four components of 
the (2,3)-quaternionic entry is set to zero.
Then, we have (falling into the same tight cluster in Fig.~\ref{fig:functs},
when the $\frac{4}{3}$-power of its
normalized form is plotted)
\begin{equation} \label{missing2}
\mathcal{S}^{Bures}_{[\hat{(2,3)}]}=
\begin{cases}
 \frac{1}{8} \pi ^2 \left(4-\sqrt{2} \log \left(3+2
   \sqrt{2}\right)\right) & \mu >1 \\
 \frac{1}{4} \pi  \left(\mu  \sqrt{1-\mu ^2}-\sin
   ^{-1}(\mu )\right) \left(\sqrt{2} \log \left(3+2
   \sqrt{2}-4\right)\right) & 0 < \mu <1
\end{cases}.
\end{equation}

We have been able, further, using the formulas of Dittmann \cite{explicit},
to compute the Bures volume elements for the 
corresponding (five-dimensional) real and 
(seven-dimensional)  complex scenarios, in which {\it both} the $\{2,3\}$ and $\{1,2\}$ entries are allowed to 
freely vary.  But these volume elements do not appear, now, 
to fully factorize into
products of functions 
(as is the case for (\ref{V23real}) and (\ref{V23complex})) 
involving just $\rho_{11}, \rho_{22}, \mu$  and just the off-diagonal
variables $x_{ij}$'s and $y_{ij}$'s. The requisite integrations are, then, 
more problematical and it seemed impossible to obtain a univariate
separability function of $\mu$.

For instance, in this regard, we have for the indicated five-dimensional real 
scenario that 
\begin{equation}
dV^{Bures}_{[(1,2),(2,3)]}=
\frac{1}{4} \sqrt{\frac{A}{B C (D +E)}} d \rho_{11} d\rho_{22} 
d x_{12} d x_{23}  d \mu,
\end{equation}
where
\begin{equation}
A=-\rho _{11}^2 \rho _{22}^2 \left(\rho _{11}+\rho
   _{22}-1\right) \left(\left(\mu ^2-1\right) \rho
   _{22}+1\right),
\end{equation}
\begin{equation}
B= \left(\rho _{22} \mu ^2+\rho _{11}\right)^2, C=x_{12}^2+x_{23}^2-1,
\end{equation}
\begin{equation}
D= \left(\rho _{11}+\rho _{22}\right) \left(x_{12}^2 \rho
   _{22} \left(\rho _{22} \mu ^2+\rho
   _{11}\right)^2-\left(\left(\mu ^2-1\right) \rho
   _{22}+1\right) \left(-\rho _{11}^2+\rho _{11}+\mu ^2
   \rho _{22}^2\right)\right)
\end{equation}
and
\begin{equation}
E=-x_{23}^2 \rho _{22} \left(\rho _{11}+\rho _{22}-1\right)
   \left(-\rho _{11}^2+\rho _{11}+\mu ^2 \rho
   _{22}^2\right).
\end{equation}
So, no desired factorization is apparent.

However, the computational situation greatly improves if we let the (1,4) 
and (2,3)-entries be the two free ones. (These entries are the specific ones
that are interchanged under 
the operation of partial transposition, so there is a greater 
evident symmetry in such a scenario.) 
Then, we found that the three Bures 
volume elements all do factorize into products of functions of
off-diagonal entries and functions of diagonal entries. We have
\begin{equation}
dV^{Bures}_{[(1,4),(2,3)]} = 
\frac{1}{8} \sqrt{-\frac{1}{\left(x_{14}^2-1\right)
   \left(x_{23}^2-1\right) \left(\rho _{22}+\rho
   _{33}-1\right) \left(\rho _{22}+\rho _{33}\right)}} d \rho_{11} 
d \rho_{22} d \rho_{33} d x_{14} d x_{23},
\end{equation}
where simply for succinctness, we now show the volume elements before 
replacing the $\rho_{33}$ variable by $\mu$. 
(We note that the expression for 
$dV^{Bures}_{[(1,4),(2,3)]}$ is independent of 
$\rho_{11}$.)
For the corresponding complex scenario,
\begin{equation}
dV^{Bures}_{[\widetilde{(1,4)},\widetilde{(2,3)}]} =
\frac{1}{8} \sqrt{\frac{F}{G}}  d \rho_{11}
d \rho_{22} d \rho_{33} d r_{14} d r_{23} d \theta_{14} d \theta_{23},
\end{equation}
where
\begin{equation}
F=-r_{14}^2 r_{23}^2 \rho _{11} \rho _{22} \rho _{33}
   \left(\rho _{11}+\rho _{22}+\rho _{33}-1\right),
\end{equation}
and
\begin{equation}
G=\left(r_{14}^2-1\right) \left(r_{23}^2-1\right) \left(\rho
   _{22}+\rho _{33}-1\right)^2 \left(\rho _{22}+\rho
   _{33}\right)^2,
\end{equation}
and we have now further shifted to polar coordinates, 
$x_{ij} + {\bf{i}}  y_{ij} = r_{ij} (\cos{\theta_{ij}} + {\bf{i}} 
\sin{\theta_{ij}})$. 
For the quaternionic scenario, we have 
(using two sets of hyperspherical coordinates $(r_{14}, \theta_{14}^{(1)}, \theta_{14}^{(2)},\theta_{14}^{(3)})$ and $(r_{23}, \theta_{23}^{(1)}, \theta_{23}^{(2)}, \theta_{23}^{(3)})$),
\begin{equation}
dV^{Bures}_{[\widehat{(1,4)},\widehat{(2,3)}]} = \frac{1}{8} \sqrt{\frac{\tilde{F}}{\tilde{G}}}  d \rho_{11}
d \rho_{22} d \rho_{33} d r_{14} d r_{23} d \theta_{14}^{(1)} d \theta_{14}^{(2)} d \theta_{14}^{(3)} d \theta_{23}^{(1)} d \theta_{23}^{(2)} d \theta_{23}^{(3)},
\end{equation}
where
\begin{equation}
\tilde{F}=\sin ^2\left(\theta _{14}^{(1)}\right) \sin \left(\theta _{14}^{(2)}\right)
   \sin ^2\left(\theta_{23}^{(1)}\right) \sin \left(\theta_{23}^{(2)}\right)
   r_{14}^3 r_{23}^3 \rho _{11}^{3/2} \rho _{22}^{3/2}
   \left(-\rho _{11}-\rho _{22}-\rho _{33}+1\right)^{3/2}
   \rho _{33}^{3/2}
\end{equation}
and
\begin{equation}
\tilde{G}=\sqrt{1-r_{14}^2} \sqrt{1-r_{23}^2} \left(\rho _{22}+\rho
   _{33}-1\right)^2 \left(\rho _{22}+\rho _{33}\right)^2.
\end{equation}

The total Bures volume for the first (real) of these three scenarios is 
$V^{Bures}_{tot/[(1,4),(2,3)]} = \frac{\pi^3}{64} \approx 0.484473$, for the second (complex) 
scenario, $V^{Bures}_{tot/[\widetilde{(1,4)},\widetilde{(2,3)}]} = 
\frac{\pi^4}{192} \approx 0.507339$, and for the third (quaternionic), 
$V^{Bures}_{tot/[\widehat{(1,4)},\widehat{(2,3)}]} = 
\frac{\pi^6}{245760} \approx 0.0039119$.

In the 
two corresponding 
Hilbert-Schmidt (real and complex) analyses 
we have previously reported, we had the results  \cite[eq. (28)]{slater833},
\begin{equation} \label{suggestion}
\mathcal{S}^{HS}_{[(1,4),(2,3)]}(\mu) = 
\begin{cases}
 4 \mu & 0\leq \mu \leq 1 \\
 \frac{4}{\mu} & \mu >1
\end{cases}.
\end{equation}
and \cite[eq. (34)]{slater833}
\begin{equation} \label{secondmixed}
\mathcal{S}^{HS}_{[\widetilde{(1,4)},\widetilde{(2,3)}]}(\mu)  =
\begin{cases}
 \pi ^2 \mu^2 & 0\leq \mu \leq 1 \\
 \frac{\pi ^2}{\mu^2 } & \mu >1
\end{cases},
\end{equation}
thus, exhibiting the indicated exact (Dyson sequence) 
proportionality relation. 
We now found, for the two Bures analogs,  that 
\begin{equation} \label{suggestion2}
\mathcal{S}^{Bures}_{[(1,4),(2,3)]}(\mu) =
\begin{cases}
 \pi ^2 & \mu =1 \\
 2 \pi  \csc ^{-1}(\mu ) & \mu >1 \\
 2 \pi  \sin ^{-1}(\mu ) & 0<\mu <1
\end{cases},
\end{equation}
\begin{equation} \label{secondmixed2}
\mathcal{S}^{Bures}_{[\widetilde{(1,4)},\widetilde{(2,3)}]}(\mu)  =
\begin{cases}
 16 \pi^2 & \mu =1 \\
 16 \pi^2 \left(1-\frac{\sqrt{\mu ^2-1}}{\mu }\right) & \mu >1
   \\
 16 \pi^2 \left(1-\sqrt{1-\mu ^2}\right) & 0<\mu <1
\end{cases},
\end{equation}
and, further still, for the quaternionic scenario,
\begin{equation} \label{thirdmixed2}
\mathcal{S}^{Bures}_{[\widehat{(1,4)},\widehat{(2,3)}]}(\mu)  =
\begin{cases}
 \frac{16 \pi ^4}{9} & \mu =1 \\
 -\frac{8 \pi ^4 \left(2 \left(\sqrt{\mu ^2-1}-\mu \right)
   \mu ^2+\sqrt{\mu ^2-1}\right)}{9 \mu ^3} & \mu >1 \\
 \frac{8}{9} \pi ^4 \left(-\sqrt{1-\mu ^2} \mu ^2-2
   \sqrt{1-\mu ^2}+2\right) & 0<\mu <1
\end{cases}.
\end{equation}
Employing these several results, we obtained that 
$V^{Bures}_{sep/[(1,4),(2,3)]} \approx 0.1473885131$,
$V^{Bures}_{sep/[\widetilde{(1,4)},\widetilde{(2,3)}]} 
\approx 0.096915844$, and
$V^{Bures}_{sep/[\widehat{(1,4)},\widehat{(2,3)}]}
\approx 0.000471134100$
giving us real, 
complex and quaternionic separability probabilities of 0.3042243652,
0.19102778 and 0.120436049.

We see that for values of $\mu \in [0,1]$, the {\it 
normalized} forms of these
three Bures separability functions are {\it identical} to the three
obtained above ((\ref{Bures1}), (\ref{Bures2}), (\ref{Bures3})) 
for the corresponding {\it single}-nonzero-entry scenarios. 
While those earlier functions were all constant for $\mu>1$, we now have
symmetrical behavior about $\mu=1$ in the form, $\mathcal{S}^{Bures}_{scenario}(\mu) =
\mathcal{S}^{Bures}_{scenario}(\frac{1}{\mu})$.

\begin{figure}
\includegraphics{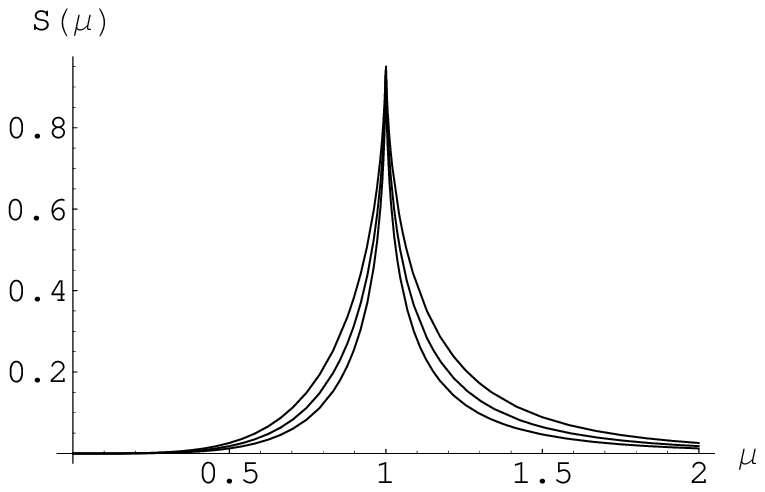}
\caption{\label{fig:functs2} Joint plot of the
normalized Bures {\it quaternionic}
separability function
$\frac{9 \mathcal{S}^{Bures}_{[\widehat{(1,4)},\widehat{(2,3)}]}(\mu)}{4}$,
the {\it square} of
the normalized Bures {\it complex} separability function
$\frac{\mathcal{S}^{Bures}_{[\widetilde{(1,4)},\widetilde{(2,3)}]}(\mu)}{16 \pi^2}$,
and the {\it fourth} power of the normalized Bures {\it real} separability
function $\frac{\mathcal{S}^{Bures}_{[(1,4),(2,3)]}(\mu)}{\pi^2}$. 
Over the interval
$\mu \in [0,1]$, the three functions are identical---with the same 
order of dominance---to those in 
Fig.~\ref{fig:functs}.}
\end{figure}

In Fig.~\ref{fig:functs2}, we show the analogous plot to Fig.~\ref{fig:functs},
using the normalized (to equal 1 at $\mu=1$) 
forms of the three  additional Bures separability functions 
((\ref{suggestion2}), (\ref{secondmixed2}), (\ref{thirdmixed2})). 
We again, of course, observe a very close fit to the type of proportionality relations
{\it exactly} observed in the Hilbert-Schmidt case 
((\ref{suggestion}), (\ref{secondmixed})).

We were, further, able to compute the Bures volume element for the
{\it three}-nonzero-entries
complex scenario $[\widetilde{(1,2)},\widetilde{(1,4)},
\widetilde{(2,3)}]$, but it was considerably more complicated in form than 
those reported above, so no 
additional analytical progress seemed possible.

Regarding the possible computation of Bures separability functions for the 9-dimensional real and 15-dimensional complex two-qubit states, we have found,
preliminarily,  
that the corresponding metric tensors (using the Bloore parameterization) 
decompose into $3 \times 3$ and
$6 \times 6$, and $3 \times 3$ and $12 \times 12$ blocks, respectively.
The $3 \times 3$ blocks themselves 
are identical in the two cases, and of precisely 
the (simple diagonal) form (if we employ hyperspherical coordinates) that 
Akhtarshenas found for the Bures metric using the coset 
parameterization \cite[eq. (23)]{iran2}. They, thus, depend only upon the 
diagonal entries (while in \cite{iran2}, the dependence, quite 
differently, was upon the 
eigenvalues). It appears, 
though, that the determinants---for which we presently lack 
succinct formulas---of the complementary
$6 \times 6$ and $12 \times 12$ blocks, do depend upon all, diagonal and
non-diagonal, parameters.

The close proximity observed in this study between 
certain separability results for the 
Hilbert-Schmidt and Bures metrics is perhaps somewhat similar in nature/explanation to a form of high similarity also observed in our previous analysis
\cite{slaterPRA}. There, large scale numerical (quasi-Monte Carlo) 
analyses strongly suggested that the ratio of Hilbert-Schmidt separability
probabilities of generic (rank-6) 
qubit-qutrit states ($6 \times 6$ density matrices) 
to the separability probabilities
of generically minimally degenerate (boundary/rank-5) 
qubit-qutrit states was equal to 2. (This has since been formally 
confirmed and generalized---in terms of 
positive-partial-transpose-ratios---to arbitrary bipartite systems
by Szarek, Bengtsson and 
{\.Z}yczkowski in \cite{sbz}. 
They found that the set of positive-partial-transpose states 
is ``pyramid decomposable'' and, hence, is a body of constant height.) 
Parallel numerical ratio estimates 
also obtained in \cite{slaterPRA} based on 
the Bures (and a number of other monotone) metrics were also surprisingly close to 2, as well (1.94334 in the Bures case \cite[Table IX]{slaterPRA}). 
However, no exact value for the Bures qubit-qutrit ratio has ever been 
established, and our separability function 
results above, might be taken to suggest that
the actual Bures ratio is not exactly equal to 2, but only quite close to it.
(Possibly, in these regards, 
the Bures metric might profitably be considered as some
perturbation of the flat Euclidean metric (cf. \cite{gross}).)

We plan to continue to study the forms the Bures separability functions take
for qubit-qubit and qubit-qutrit scenarios, with the hope that we can achieve
as much insight into the nature of Bures separability probabilities,
if not more, than we obtained by examining the analogous
Hilbert-Schmidt separability functions \cite{slater833}.
(In \cite{slaterJGP}, we had formulated, based on extensive numerical
evidence, conjectures---involving the silver mean, $\sqrt{2}-1$---for the
Bures [and other monotone metric]
separability probabilities of the 15-dimensional convex set of
[complex] qubit-qubit states, which we would further aspire to test. 
One may also consider the use of monotone metrics other than the 
{\it minimal} Bures one 
\cite{andai}---such as the Kubo-Mori and Wigner-Yanase.)
The analytical challenges to further progress,
however, appear quite formidable.

\begin{acknowledgments}
I would like to express gratitude to the Kavli Institute for Theoretical
Physics (KITP)
for computational support in this research.

\end{acknowledgments}

\bibliography{Truncated}

\end{document}